\documentclass[12pt]{article}
\input{epsf}
\usepackage{epsf}

\def\AJ{{Astroph. J.} }
\def\AJL{{Ap. J. Lett.} }

\def\ASJ{{Astron. J.} }

\def\MNRAS{{Mon. Not. R. Ast. Soc.} }

\def\NP{{Nucl. Phys.} }
\def\PL{{Phys. Lett.} }
\def\PR{{Phys. Rev.} }
\def\PRL{{Phys. Rev. Lett.} }
\def\PRTS{{Physics Reports} }

\def\lsim{\mathrel{\rlap{\lower4pt\hbox{\hskip1pt$\sim$}}
    \raise1pt\hbox{$<$}}}
\def\gsim{\mathrel{\rlap{\lower4pt\hbox{\hskip1pt$\sim$}}
    \raise1pt\hbox{$>$}}}

\def\beq{\begin{equation}}
\def\eeq{\end{equation}}
\def\beqa{\begin{eqnarray}} 
\def\eeqa{\end{eqnarray}}

\def\laq{\raise 0.4 ex \hbox{$<$}\kern -0.8 em\lower 0.62 ex\hbox{$\sim$}}
\def\gaq{\raise 0.4 ex \hbox{$>$}\kern -0.7 em\lower 0.62 ex\hbox{$\sim$}}

\newcommand\vc[1]{\ensuremath{\mathbf{#1}}}

\newcommand\av[1]{\bar{#1}}
\newcommand\req[1]{Eq.~\ref{#1}}

\begin{document}

\begin{center}

{\Large \bf Generalized Chaplygin Gas in a modified gravity approach}

\vspace{1cm}
T. Barreiro\footnote{Email address: tiagobarreiro@fisica.ist.utl.pt}
and A. A. Sen\footnote{Email address: anjan@cfif3.ist.utl.pt} 

\vspace{0.5cm}
{ Departamento de F\'\i sica, Instituto Superior T\'ecnico \\
Av. Rovisco Pais 1, 1049-001 Lisboa, Portugal}

\vspace{0.5cm}
{\bf Abstract}

\end{center}
\noindent

We study the generalized Chaplygin gas (GCG) scenario in a modified gravity
approach.
That is, we impose that our universe has a pure dust configuration, and allow for a modification of gravity that yields a GCG specific scale factor evolution.
Moreover, assuming that this new hypothetical gravity theory obeys a generalization of Birkhoff's law, we determine the Schwarzschild-like metric in this new modified gravity.
We also study the large scale structure formation in this model. Both the linear and non-linear  growth are studied together with the growth of the velocity fluctuation in the linear perturbation theory. We compare our results with those corresponding to the $\Lambda$CDM model and discuss possible distinguishable features.

\newpage 
\section{Introduction}

The nature of the dark energy (DE) that is responsible for today's cosmic acceleration is an open and tantalizing mystery \cite{SN}. It leaves room for novel theoretical explanations and new cosmological scenarios. The most obvious is the $\Lambda$ Cold Dark Matter model ($\Lambda$CDM), comprising dark matter and a non zero cosmological constant-$\Lambda$. But the difficulty of explaining the value of this $\Lambda$ term from fundamental physics leaves motivations for other phenomenological proposals. Two such alternative avenues for DE model building in which the DE is dynamical are the quintessence \cite{QT} and k-essence \cite{KS} models. Similar to inflation, the first one is constructed with an ordinary  minimally coupled scalar field whose equation of state is a function of time. Amongst them, the tracker quintessence models \cite{TR} have the advantage of allowing the current accelerating epoch to be reached from a large set of initial conditions.

Recently, there are suggestions that the present acceleration of the universe is not due to any new unknown component in the cosmic soup, but due to a modification of the gravitational physics at scales typically much smaller than today's horizon. A few suggestions are particularly interesting in this regard: the Dvali-Gabadadze-Poratti (DGP) brane induced gravity model \cite{DGP}, the Cardassian model proposed by Freese and Lewis\cite{FL} and also the recent model by Dvali and Turner \cite{DT}. All these models lead to late-time acceleration with a modified Friedman equation and with no explicit dark energy component. The basic feature of this approach is that the geodesics in a static spherically symmetric spacetime may completely determine the cosmological evolution, i.e., the present acceleration of the universe is not driven by any extra dark energy component but by the matter content itself. Given a complete cosmological evolution of our universe as suggested by observ!
 ations (i.e. given a specified scale factor $a(t)$), if we impose that it is due to a pure dust configuration, we can study what modification of the gravity is necessary to yield such a scale factor. Moreover, imposing that this new hypothetical gravity theory obeys a generalization of Birkhoff's law, one can get a unique modification. This procedure of determining the Schwarzschild-like metric of the new modified gravity that gives the prescribed cosmological evolution with a dust-filled universe has been studied extensively by Lue et.al \cite{LUE}.
Also, in a recent paper Multamaki et.al \cite{MUL} have studied the growth of large scale structure in the DGP and Cardassian model. They put forward a general formalism for calculating the growth of both linear and non-linear fluctuations in models with non-standard Friedman's equations.

Another alternative to the quintessence model which has attracted great interest in recent times is the so-called generalized Chaplygin gas (GCG) model \cite{GCG}. The model explains the acceleration of the universe via an exotic equation of state resulting in a behaviour like dark matter at early times and dark energy at late times.  The GCG is characterized by the equation of state 

\beq
p_{ch} = - {A \over \rho_{ch}^\alpha},
\eeq
where $A$ and $\alpha$ are positive constant. For $\alpha=1$, the equation of state is 
reduced to so-called Chaplygin gas scenario \cite{KAMEN}.

Inserting the above equation of state in the 
energy conservation equation, one can integrate it to obtain
\beq
\rho_{ch} = \left(A + {B \over a^{3(1+\alpha)}}\right)^{1/(1+\alpha)},
\eeq
where $B$ is an arbitrary constant of integration which should also be positive. One can see at once that this energy density interpolates between a dust like configuration in the past and a de-Sitter like one in the late times. This property makes the GCG model an interesting candidate for the unification of dark matter and dark energy.

Using the above expression for $\rho$ in the Einstein equation for $H(t)$, one gets

\beq\label{hub1}
H^{2}(t) = {8\pi G \over 3} \left(A + {B \over a^{3(1+\alpha)} (t)}\right)^{1/(1+\alpha)}.
\eeq
This expression for $H(t)$ is a very good fit for the observational data as far the background cosmology is concerned. This has been successfully confronted with various observational test: high precision Cosmic Microwave Background data \cite{CMB}, Supernova data \cite{GSN} and data from gravitational lensing \cite{GL}.

But despite all these pleasing features, the main difficulty with such an unified model  is that it produces unphysical oscillations or exponential blow-up in the matter power spectrum at present \cite{SAND}.  
Moreover, it was shown that the linear approximation breaks down at an early stage, implying that a more careful approach, including nonlinear effects, should be taken into account \cite{AVEL}.
Some efforts have been made to circumvent the previous problem by adding the baryons into the model which is not accounted for by GCG \cite{BECA}. 
GCG has also been treated as a dark energy model in combination with dark matter, where only the dark matter part is perturbed \cite{MUL2}.
In a very recent work, it has been shown that this equation of state represents uniquely an interacting mixture of decaying dark matter and a cosmological constant once one excludes the possibility of having a phantom like dark energy and in such a scenario one can avoid the problem of having unphysical features in the matter power spectrum \cite{REV}.

In this letter, we are considering the GCG in a  modified gravity approach as described by Lue et.al \cite{LUE}. That is, we assume that the background evolution is due to some kind of modified gravity, rather than an energy density with a GCG equation of state. We impose the same background evolution, but with an energy density consisting purely of matter (we ignore the residual radiation term at present).
We deduce the corresponding Schwarzschild-like metric for this new modified gravity. We also study both the linear and second order density perturbations in this modified GCG universe and compare our result with that of a standard $\Lambda$CDM model.

\section{Modified GCG model}

We assume that our background universe is well described by the $H(t)$ given by \req{hub1} but that it  only contains a pure matter 
configuration. That is, our assumption is that the only energy density in the universe has an equation of state $w = 0$ with a conservation equation
\begin{equation}\label{matter}
\dot{\rho} = -3 H \rho
\end{equation} 
Subsequently we look for the possible modification of Einstein gravity which results in
the above $H(t)$. Noting that for the dust in \req{matter}, $\rho(t) \propto a^{-3}(t)$, one can see that such 
modification is given by

\beq\label{hubble}
H^{2}(t) = {8\pi G \over 3} \left(A + \rho^{\alpha+1}\right)^{1/(1+\alpha)}.
\eeq
One can relate the constant $A$ with the matter density parameter $\Omega_m$ as
$A = (3H_{0}^{2}/8 \pi G)^{(\alpha+1)}(1 - \Omega_{m}^{\alpha+1})$.
Also, we have identified the previous constant $B$ with $\rho_0^{\alpha +1}$ ($\rho_0$ being the present matter energy density), which can be done with no loss of generality. 
Notice that since we asume our energy density restricted to dust, we have $\Omega_{T} = \Omega_{m} < 1$. Alternatively, we can interpret the constant $A$ as a cosmological constant, adding an extra component to the energy density, like what is done in \cite{DGP}. Either way, the evolution equation for the matter density \req{matter} is unchanged and does not couple to any other fluid. This is in contrast to what was done in \cite{REV}.
(Furthermore, when $\alpha = 0$ this expression reduces to a $\Lambda$CDM model. Since we are only considering matter as our energy density, this means that the two models will have exactly the same evolution. That is, for $\alpha = 0$ we have a ``modified gravity'' $\Lambda$CDM model).
\footnote{For the usual GCG scenario, the $\alpha = 0$ case is also equivalent to a $\Lambda$CDM model, however one has to be careful with the background used to calculate the perturbations when comparing between the two \cite{AVEL2}.}
Using \req{hubble}, one can write the modified Einstein equation as

\beq\label{h2}
H^{2} = H_{0}^{2} \; g(x),
\eeq
where $g(x) = \left[(1-\Omega_{m}^{\alpha+1})+x^{\alpha+1}\right]^{1/\alpha+1}$  and $x$ is a dimensionless quantity defined as $x = {8 \pi G \over{3}}\; \rho/H_0^2$. For $x>>1$, $g(x) 
\rightarrow x$ and one recovers the standard Einstein equation in the early universe, whereas 
in the late time the gravity is modified.
Since we are assuming the scale factor evolution to be due to a modification of Einstein's gravity, it is interesting to consider what effects this modification could have at astrophysical scales.
This is particularly easy to do if we impose an extra condition on the theory,
namely that it obeys a generalization of Birkhoff's law.
Following the procedure in \cite{LUE}, it is possible to
deduce the Schwarzschild-like 
metric of this modified gravity around a spherically symmetric matter source. In general, this is given by

\beq\label{swarz}
g_{00} = g_{rr}^{-1} = 1 - r^{2} H_{0}^{2}g(\frac{r_{c}^{3}}{r^{3}}),
\eeq
where $g$ is the function defined after \req{h2}, 
$r$ is the usual radial coordinate, and $r_c$ is a measure of the distance scale over 
which gravity is modified. It is given by $r_{c} = ({2GM\over{H_0}^2})^{1/3}$,  $M$ being the  mass of the spherically symmetric gravitating object of energy density $\rho$ and radius $r$ and is given by $M = {4\pi\over{3}}\rho(t)r^3$. 
Since $g(x) \to x$ when $x \gg 1$,
and we have $x = {r_c^3\over{r^3}}$,
it easy to check that the metric will go to the usual Schwarzchild metric at small $r \ll r_c$.
However, for values of $r \gg r_c$ the metric will be modified, and one then expects to see a significant deviation from the usual Einstein gravity. As an example, consider the universe within the horizon, with a mass $M \approx \frac{1}{G H_0}$. This will give us a scale $r_c \approx H_0^{-1}$, as expected, since it is approximately the scale at which acceleration sets in.

Even if we consider radii smaller than $r_c$, one can still  try to use 
this metric, with its small modifications, to
impose some astrophysical bounds on this model. 
As a particular example, for the solar system
the orbits of the planets are well within the value of $r_c$.
However, we can expand the metric in powers of $\frac{r}{r_c}$ and check what the first order correction to Einstein's gravity gives.
This was done in \cite{LUE}, where the authors obtained a formula
for the precession of the perihelion. For our modified gravity, the correction will be proportional to $\left( r/r_c\right)^{\frac{3}{2}(2 \alpha + 1)}$. If the exponent is negative, the correction would be huge for very small values of $r$. This  imposes a bound on our parameter space
of $\alpha > -\frac{1}{2}$.
In our numerical simulations we will restrict ourselves to $\alpha \geq 0$.

\section{The perturbed equation}

To study the density perturbations, 
we consider the evolution of a perturbed ideal fluid, with a shear free four-velocity $u^\mu$. We choose a coordinate system such that $u^{\mu} = (1, \dot{a} \vc{x} + \vc{v})$, with $\vc{v}$ being the peculiar velocity of the fluid.

We define the perturbed part of the energy density $\delta$ as  
\beq
\rho = \av{\rho} \, (1 + \delta)
\eeq
where $\av{\rho}$ is the background energy density (and for the rest of this paper
barred quantities refer to their background values). The perturbed part of the continuity equation is then given by \cite{PEEB}
\beq
\frac{d \delta}{d \tau} + (1+\delta) \theta = 0
\eeq
where $\tau$ is the conformal time, $dt = a d \tau$, and $\theta = \vc{\nabla} \cdot \vc{v}$.

We can get a second equation for the fluid evolution from Raychaudhuri's equation. This can be written as \cite{MUL}
\beq
\frac{\dot{\theta}}{a} + \frac{\theta}{a} \av{H} = 3 (\dot{H} + H^2) - 3 (\dot{\av{H}} + \av{H}^2) \;.
\eeq

Combining the two, we get an equation for the perturbed energy density evolution, for a general matter density background,
\beq
\frac{d^2 \delta}{d \eta^2}
+ (2 + \frac{\dot{\av{H}}}{\av{H}^2} ) \frac{d \delta}{d \eta}
- \frac{4}{3  (1+\delta)} \left( \frac{d \delta}{d \eta} \right)^2 
= 
-3 \frac{1 + \delta}{\av{H}^2} \left( (\dot{H} + H^2) - (\dot{\av{H}} + \av{H}^2) \right) \;
\eeq

Following \cite{MUL}, we perform an expansion in the perturbation $\delta$
\beq
\delta = \sum_{i=1}^{\infty} \delta_i = \sum_{i=1}^{\infty} \frac{D_{i}(\eta)}{i !} \delta_{0}^{i}
\eeq
where $\delta_0$ is a series expansion parameter. We will be interested in the first two terms in the expansion, for which the linearized equations are, in general,
\beqa
&  & D_{1}'' + (2 + \frac{\dot{\av{H}}}{\av{H}^2}) D_{1}' + 3 c_{1} D_{1} = 0
\\
& & D_{2}'' + (2 + \frac{\dot{\av{H}}}{\av{H}^2}) D_{2}'
- \frac{8}{3} (D_{1}')^2 + 3 c_1 D_2 + 6 (c_1 + c_2) D_1^2 = 0
\eeqa

\begin{figure}
\begin{center}
\epsfxsize 3.5in
\epsfbox{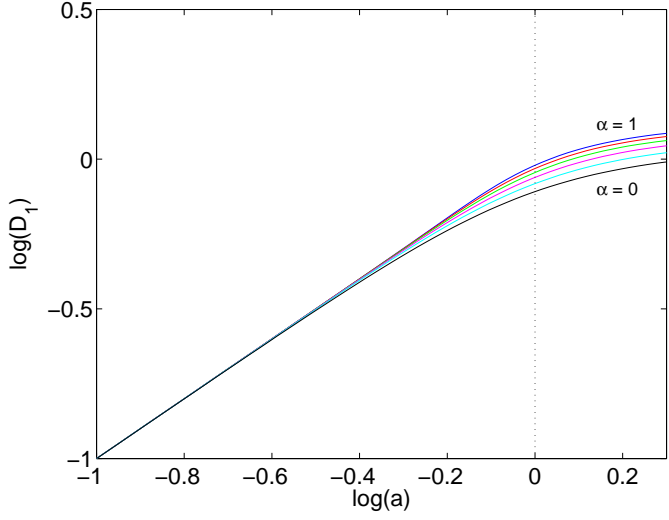}
\caption{Linear growth $D_{1}$ for $\alpha = 0, 0.2, 0.4, \ldots, 1.0$ from bottom to top.
}
\end{center}
\end{figure}

It is straightforward to compute the two terms $c_1$ and $c_2$ for the ``generalized Chaplygin'' scenario. They come out as

\beqa
c_1 & = & -\frac{\big[ (1 + 3 \alpha) A + \rho^{\alpha+1}\big] \;\rho^{\alpha+1} }{2 (A + \rho^{\alpha+1})^2} \\
c_2 & = & \frac{\alpha A \rho^{\alpha+1}}{4 (A + \rho^{\alpha+1})^3} \big[ -(1 + 3 \alpha) A + (2 + 3 \alpha) \rho^{\alpha + 1}\big]
\eeqa
and it is easy to check that they reduce to the standard cosmological constant result for $\alpha = 0$, as expected. The second order perturbation is related with the skewness of the density field at large scale. The q-order moment of the fluctuating field is related with the perturbation as
\beq
m_q = <\delta^q>.
\eeq
The normalized skewness is given by
\beq
S_3 = {m_3\over{m_2}}
\eeq
which can be written in terms of the first and second order perturbations. For Gaussian perturbations $<\delta_{1}^{3}> = 0$, so that one gets

\beq
S_3 = 3 {D_2\over{D_1}}
\eeq
 In the standard CDM model, this coefficient can be calculated exactly to give $S_{3} = 34/7 \approx 4.86$.

\begin{figure}
\begin{center}
\epsfxsize 3.5in
\epsfbox{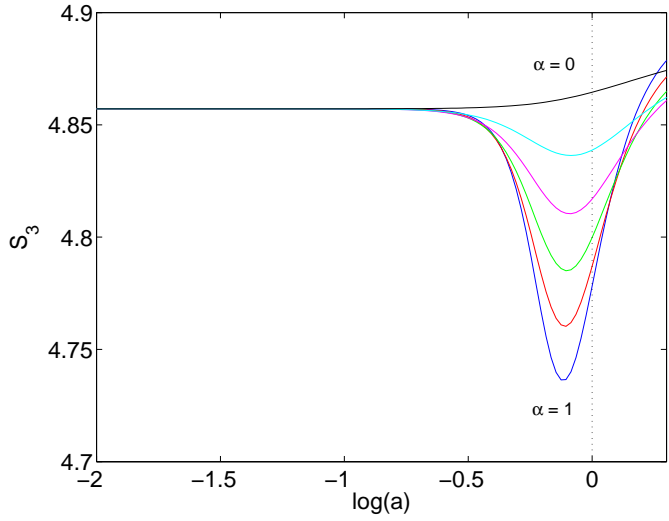}
\caption{Non-linear growth $S_{3}$ for $\alpha = 0, 0.2, 0.4, \ldots, 1.0$ from top to bottom (at present).
}
\end{center}
\end{figure}

\subsection*{Numerical results}

At early times, there is no difference in the evolution between the standard CDM 
and any of the proposed modified gravity scenarios, including ours. Therefore, for our numerical results, we start the simulation at an early time, $a = 10^{-3}$, taking as initial conditions the standard CDM solution, $\delta \propto a$. In calculating the second order perturbation, initial conditions are chosen such that the standard solution $S_{3} = 34/7$ is valid from the beginning. In figure 1, we have plotted the first order perturbation $D_{1}$ for different values of $\alpha$. In Figure 2, we have shown that skewness $S_{3}$ for the second order perturbation. In plotting these figures, we have assumed a fixed $\Omega_{m} = 0.3$, which sets the value of $A$ for each $\alpha$ (cf. the comment after \req{hubble}).

\begin{figure}
\begin{center}
\epsfxsize 3.5in
\epsfbox{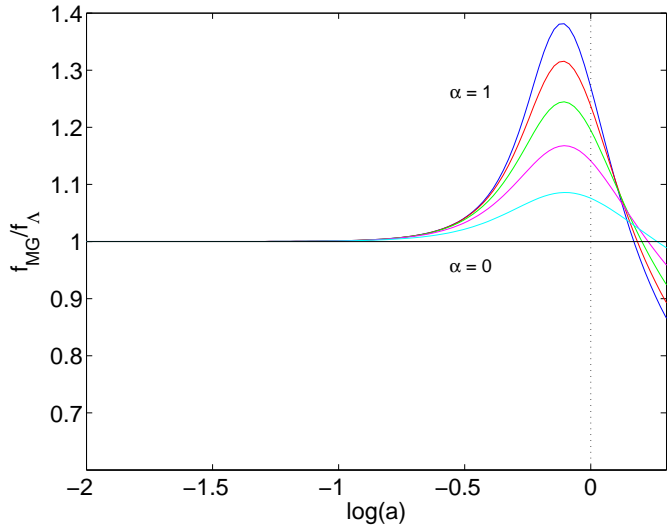}
\caption{Variation of the ratio $f_{MG}/f_{\Lambda}$ for $\alpha = 
0,0.2, 0.4, \ldots, 1.0$ from bottom to top (at present).}
\end{center}
\end{figure}

The general form of linear growth $D_{1}$ is similar to the $\Lambda$CDM case but one can have more linear structure formed at present with increasing values of $\alpha$. This will also result in a higher $\sigma_{8}$ value (the rms fluctuation on a sphere of 8 Mpc/h) at present than the corresponding $\Lambda$CDM value.  Also one does not have any oscillations or exponential blow in the matter power spectrum at present as in the usual GCG scenarios. This is expected as a consequence of having only a dust configuration in the universe with a vanishing pressure. 

In contrast,  the second order perturbation or $S_{3}$ evolves in a completely different way from the $\Lambda$CDM model, as shown in figure 2. However the variations from the $\Lambda$CDM case at present are approximately $2\%$ for $\alpha$ as high as 1. Current estimates for $S_3$ agree with the standard prediction but with large uncertainties, of the order of $20\% - 30\%$ \cite{BERN}. Hence the current observational result can not be used to differentiate between this modified GCG gravity with the standard $\Lambda$CDM model using $S_3$. Although the Sloan Digital Sky Survey (SDSS) is  expected to reduce these uncertainties to around $5\%$, still this will not be enough \cite{GAZ}. We have also checked that varying $\Omega_{m}$ within the range $0.2<\Omega_{m}<0.4
$ does not change the result much.

\begin{figure}
\begin{center}
\epsfxsize 3.5in
\epsfbox{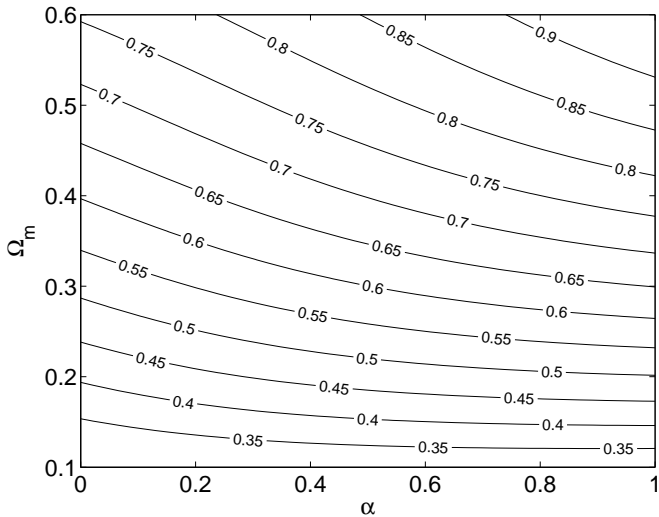}
\caption{Contours for the present day value of $f$ in $\alpha$-$\Omega_{m}$ plane.
}
\end{center}
\end{figure}

In figure 3, we have shown the variation of the ratio $f_{MG}/f_{\Lambda}$, 
where $f \equiv$ d $\ln D_{1}/$d $\ln a$. One can see from figure 3 that  for $\alpha = 1$,
 $f$ can increase up to $40\%
 $ from the corresponding $\Lambda$CDM value. Notice that 
$f$ governs the growth of the velocity fluctuations in the linear perturbation theory;
 therefore the large deviations of $f$ with changing $\alpha$ is detectable via precision
 measurements of large scale structure through joint measurements of the redshift-space
 power spectrum anisotropy and bi-spectrum from $z=0$ to $z\approx 2$. The SDSS should 
be able to probe this quantity with statistical error of order of a few percent \cite{GAZ}. In figure 4
 we have also shown the contour plot for the present value of $f$ in the $\alpha$-$\Omega_m$ 
plane. It shows that with increasing $\Omega_{m}$, the $f$ depends more strongly on $\alpha$.

\section{Conclusion}

Using as motivation the good observational fit of a GCG model to the background evolution of the universe, we have studied it in a modified gravity approach as described by Lue et.al \cite{LUE}.
We have also assumed  
 that the new gravitational physics obeys a generalization of the Birkhoff's law. This means that an observer in the gravitational field of a spherically symmetric source of mass $M$ experiences a significant deviation from the usual Schwarzschild metric at a distance scale greater than approximately $({2 G M \over{H_{0}^2}})^{1/3}$, where $H_{0}$ is the present Hubble radius. 
>From this, a simple calculation with the metric \req{swarz} shows that astrophysical bounds impose a value of $\alpha > -\frac{1}{2}$.
An interesting check to the validity of this model would entail a more detailed study of its behaviour at astrophysical scales.

We have studied the first and second order density perturbation in this model. Our results show that the linear perturbation $D_{1}$ evolves in a similar way to a $\Lambda$CDM model,  but with a larger value at present. This 
results in an enhancement in the corresponding $\sigma_{8}$ value, namely up to $22\%
$ at present for $\alpha = 1$.
In theory, the nonlinear perturbation $S_{3}$ gives a more radical signature, evolving quite differently from the $\Lambda$CDM case. However, the effect is very small, and with the present observational uncertainty, one can still not use this to distinguish between the two models. 
We have also studied the parameter $f$ related with the velocity fluctuations in the linear perturbation, and have shown that the variation obtained in comparison to a $\Lambda$CDM model is quite significant. This leads to a possible detection by the present day observations.

As expected, by considering a modified gravity approach to a GCG, we can avoid
the problem of having unwanted oscillations or exponential blow-up in the matter power spectrum at present,  as one expects in its unified dark matter-dark energy approach.

Finally, it remains to be seen how one can ultimately obtain such modifications in gravity from a fundamental theory. But given the fact that the GCG type equation of state can arise from Born-Infeld type lagrangian \cite{GCG}, one may expect that D-brane physics can shed possible light to tackle this problem. 

\section*{Acknowledgments}

\noindent
The work of A.A. Sen is financed by FCT grant SFRH/BPD/12365/2003, and T. Barreiro by FCT grant SFRH/BPD/3512/2000.


\end{document}